\documentclass[fleqn,twoside]{article}
\usepackage{rotating}
\usepackage{espcrc2}
\usepackage{amssymb}
\usepackage{flafter}

\def\ss{\scriptstyle}

\usepackage{graphicx}
\usepackage{psfig}

\newcommand{\be}{\begin{equation}}
\newcommand{\ee}{\end{equation}}
\newcommand*{\bea}{\begin{eqnarray}}
\newcommand*{\eea}{\end{eqnarray}}
\providecommand*{\ler}{\stackrel{\scriptstyle <}{\scriptstyle \sim}}
\providecommand*{\ger}{\stackrel{\scriptstyle >}{\scriptstyle \sim}}
\newcommand{\nn}{\nonumber}
\newcommand{\Frac}[2]{\frac{\displaystyle{#1}}{\displaystyle{#2}}}
\newcommand{\lsim}{\raise0.3ex\hbox{$\;<$\kern-0.75em\raise-1.1ex\hbox{$\sim\;$}}}
\newcommand{\gsim}{\raise0.3ex\hbox{$\;>$\kern-0.75em\raise-1.1ex\hbox{$\sim\;$}}}

\begin{document}
\title{SUSY Seesaw and FCNC}
\author{Antonio Masiero\address[am]{Dip. di Fisica `G. Galilei', 
Univ. di Padova  \\and  INFN, Sezione di Padova, via F. Marzolo 8,  
I-35131, Padua, Italy.}, Sudhir K. Vempati \addressmark[am] and 
O.Vives\address{Theory Group, Physics Department, CERN, CH-1211,\\ 
Geneva 23, Switzerland.}}
\begin{abstract}
After a quarter of century of intense search for new physics beyond the
Standard Model (SM), two ideas stand out to naturally cope with (i) small
neutrino masses and (ii) a light higgs boson : Seesaw and SUSY. The
combination of these two ideas, i.e. SUSY seesaw exhibits a potentially
striking signature: a strong (or even very strong) enhancement of lepton
flavour violation (LFV), which on the contrary remains unobservable in the
SM seesaw.  Indeed, even when supersymmetry breaking is completely flavour 
blind, Renormalisation Group running effects are expected to generate large 
lepton flavour violating entries at the weak scale. 
In Grand Unified theories, these effects can be felt even
in hadronic physics. We explicitly show that in a class of SUSY SO(10)
GUTs there exist cases where LFV and CP violation in B-physics can constitute
a major road in simultaneously confirming the ideas of Seesaw and low-energy
SUSY. 
\end{abstract}

\maketitle

\section{Introduction}
In the Standard Model (SM) with massless neutrinos the 
three Lepton Flavour (LF) numbers are exactly conserved 
( at any order in perturbation theory). The introduction of a 
mass for the neutrinos leads to LF violation (LFV) analogously to 
the violation of flavour numbers ( strangeness, charm, etc.) in the 
quark sector. However, given that LFV in the SM has to be proportional 
to the neutrino masses, we conclude that within the SM we expect any 
LFV process other than neutrino oscillations  to be affected by 
suppression factors proportional to some power of the ratio of neutrino 
mass to the W mass. Hence, although the discovery that neutrinos are massive 
entails that LF numbers are no longer conserved in the SM, we can safely 
state that, as long as the SM represents the correct physical description, 
no LFV process like $\mu \to e+\gamma$ should ever be observed. 

The situation radically changes when we move from the SM to its 
supersymmetric (SUSY) extensions. The main difference lies in the fact 
that now we have also the scalar partners of the leptons ( sleptons) 
which carry LF numbers and hence, a priori, one may expect that there 
are contributions to LFV processes where ratios of flavour off-diagonal 
slepton masses to some average SUSY mass appear which may easily be orders 
of magnitude larger than the ratio $m_\nu /\ M_W$.

This is indeed the case. The reason for a conspicuous value of the above 
mentioned LFV entries in the slepton mass matrices is twofold. In SUSY 
extensions of the SM where the terms which break SUSY softly are not flavour 
universal, one could even imagine the off-diagonal LFV entries to be of 
the same order as the flavour conserving diagonal entries. This would be 
disastrous just because LFV would become too large ( and the same would 
happen also in the hadronic sector if flavour non-universality in the squarks 
is maximal). However, even assuming the opposite case, namely exact 
flavour universality of the soft breaking terms ( at the superlarge scale 
at which they appear in a supergravity framework) , there exists a 
remarkable property of the RG running of the slepton masses which may 
yield sizeable off-diagonal slepton masses at the scale of interest for 
our experiments, i.e. the electroweak scale. This occurs whenever the lepton
superfields have new large Yukawa couplings. The seesaw \cite{seesaw1,seesaw2} 
mechanism represents a typical context where this can be implemented. 
This was first pointed out in the SUSY seesaw model in the work of 
Borzumati and Masiero in 1986 \cite{fbam} (as I said in the talk, this 
work was prompted by some discussion that I previously had with Marciano 
and Sanda on the issue of LFV in SUSY). At 
that time we individuated the two quantities which crucially determine the 
size of the RG-induced FV off-diagonal slepton mass matrix entries: the 
Yukawa couplings responsible for the Dirac entries of the neutrino mass 
matrices and the rotating matrix establishing the mismatch in the 
diagonalisation of the lepton and slepton mass matrices. Unfortunately 
in 1986, still little was known about neutrino masses and mixings ( 
indeed, to be sure, not even the fact that neutrinos were massive and 
mixed was established!). 

The enormous amount of literature dealing with LFV in SUSY seesaw after the 
discovery of neutrino oscillations can be easily understood. Although, 
honestly, from the experimental data we find neither the Dirac neutrino 
Yukawa couplings nor the mentioned mixing matrix, it is true that all the 
information we have collected in recent years on neutrino masses and mixings 
provides important clues on the above quantities relevant in 
SUSY seesaw. Complementarily, various experiments have improved the limits
on the rare LFV decay processes over the years and in the near future, 
they are expected to do furthermore. To have an idea where we stand, here
we provide a list of present and upcoming experimental limits:
\noindent\textit{Present limits}:\\
\begin{center}
\begin{tabular}{rclc}
$BR(\mu \to e \gamma)$ &$\leq$& $1.2\times 10^{-11}$ & \cite{mega}\\
$BR(\tau \to \mu \gamma)$ &$\leq$& $3.1\times 10^{-7}$ & \cite{belletmg}\\
$BR(\tau \to e \gamma)$ &$\leq$& $3.7\times 10^{-7}$ & \cite{belletalk}
\end{tabular}
\end{center}

\noindent\textit{Upcoming limits}:\\
\begin{center}
\begin{tabular}{rclc}
$BR(\mu \to e \gamma)$ &$\leq$& $10^{-13} \div 10^{-14}$ & \cite{psi}\\
$BR(\tau \to \mu \gamma)$ &$\leq$& $10^{-8}$ & \cite{belletalk}\\
$BR(\tau \to e \gamma)$ &$\leq$& $10^{-8}$ & \cite{belletalk}
\end{tabular}
\end{center}
In this talk we are going to provide an example of how the 
interplay between new experimental data on these decays 
and theoretical progress may help in shedding light on the quantitative 
predictions on LFV in SUSY seesaw in general, and, more specifically, 
in the context of the SUSY $SO(10)$ scheme. 
Other LFV processes like $\mu \to e$ conversion in Nuclei \cite{okada},  
(Higgs mediated ) $\tau \to 3 \mu$ \cite{t3mu}, flavour violating 
Z-decays \cite{illana}, Higgs decays\cite{andrea} and other 
collider processes \cite{ruckl} are also
being investigated in the literature. 

\section{Supersymmetric Seesaw and Leptonic Flavour Violation}

The seesaw mechanism  can be incorporated in the  
Minimal Supersymmetric Standard Model in the similar manner as in 
the Standard Model, by adding right handed neutrino superfields to the
MSSM superpotential:
\bea
\label{eq1}
W &=&  W_{Y_Q} + h^e_{ij} L_i e^c_j H_1 + h^\nu_{ij} L_i \nu^c_j H_2 \nonumber\\
&+& M_{R_{ij}} \nu_i^c \nu_j^c, 
\eea
where the leptonic part has been detailed, while the quark Yukawa
couplings and the $\mu$ parameter are contained in $W_{Y_Q}$. $i,j$ are
generation indices.  $M_{R}$ represents the (heavy) Majorana mass matrix 
for the right-handed neutrinos.  Eq.(\ref{eq1}) leads to the standard seesaw 
formula for the (light) neutrino mass matrix 
\begin{equation}
\label{seesaweq}
{\mathcal M}_\nu = - h^\nu M_{R}^{-1} h^{\nu~T} v_2^2,
\end{equation}
where $v_2$ is the vacuum expectation value (VEV) of the up-type
Higgs field, $H_2$. Under suitable conditions on $h^\nu$ and $M_R$,
the correct mass splittings and  mixing angles in $\mathcal{M}_\nu$ 
can be obtained. Detailed analyses deriving these conditions are 
already present in the literature \cite{seesawreviews}.  

The above lagrangian has to be supplemented by a part containing 
supersymmetry breaking soft terms. The flavour structure of these
terms would depend on the mechanism which breaks supersymmetry and
conveys it to the observable sector. However, the accumulating 
concordance between the Standard Model (SM) expectations and the
vast range of FCNC and CP violation \cite{fcncreview} point out 
to a SUSY breaking mechanism which is flavour blind, as in mSUGRA,
Gauge-Mediation (GMSB) and Anomaly Mediation (AMSB) and its variants. 
The main observation
of \cite{fbam} was that in spite of possible flavour-blindness of 
SUSY breaking, the supersymmetrization of the seesaw leads to 
new sources of LFV\footnote{ Of the above mentioned SUSY breaking
mechanisms, this is always true in a gravity mediated supersymmetry
breaking model, but, applies also to other mechanisms under
some specific conditions \cite{yanagidagmsb,murayamaamsb}.}.
 This occurs because the flavour-blindness
of the slepton mass matrices is no longer invariant under RG evolution
from the large SUSY breaking scale down to the electroweak (seesaw) scale in the
presence of the new (seesaw) couplings \cite{hkr}.

The amount of lepton flavour violation generated by the SUSY seesaw at
the weak scale crucially depends on the flavour structure of $h^\nu$ 
and $M_R$, shown in the eq.(\ref{eq1}), the `new' sources of flavour 
violation not present in the MSSM. To see this, one has to solve 
the Renormalisation Group Equations (RGE) for the slepton mass matrices 
from the high scale to the scale of the right handed neutrinos. Below 
this scale, the running of the FV slepton mass terms is RG-invariant 
as the right handed neutrinos decouple from the theory. For the purpose of 
our illustration, a leading log estimate can easily be obtained 
for these equations. Assuming the flavour blind mSUGRA specified by 
the high-scale parameters: $m_0$, the common scalar mass, $A_0$, 
the common trilinear coupling and $M_{1/2}$, the universal gaugino mass, 
the flavour violating entries in these mass matrices at the weak scale 
are given as:

\bea
\label{rgemi}
(m_{\tilde L}^2)_{ij (i \neq j)} &\approx& -{3 m_0^2+A_0^2 \over 8 \pi^2}
\sum_k (h^\nu_{ik} h^{\nu *}_{jk}) \ln{M_{GUT} \over M_{R_k} } \nonumber \\
\eea
where $M_R$ is the scale of the right handed neutrinos. Given this,  
the branching ratios for LFV rare decays, $l_j \to l_i, \gamma$ are 
roughly estimated as \cite{fbam,largetanbeta,gabbiani,hisano} : 
\be
\label{BR}
\mbox{BR} (l_j \to l_i \gamma) \approx \Frac{ \alpha^3 ~
(~[m_{\tilde{\scriptscriptstyle{L}}}^2]_{ij} )^2 }
{G_F^2~ m^8_{SUSY}} \tan^2 \beta,
\ee
where $m_{SUSY}$ represents the typical soft supersymmetric
breaking mass, determined by $m_0, M_{1/2},$ etc., at the 
weak scale. 

From above it is obvious that if either the neutrino Yukawa couplings
or the flavour mixings present in $h^\nu$ are very tiny, the strength
of LFV will be significantly reduced. Further, if the right handed 
neutrino masses are heavier than the supersymmetry breaking scale 
(as in GMSB models), these effects would vanish. 

However, to make a more quantitative analysis of LFV in susy seesaw 
models, say, in terms of the supersymmetry breaking parameters, one 
needs to make further assumptions on the seesaw couplings of the model. 
This is because despite the huge successes we had in the neutrino physics, 
information from neutrino masses is nonetheless not sufficient to 
determine all the seesaw parameters \cite{casasibarra} in 
eq.(\ref{seesaweq}), which are crucial to compute the 
relevant LFV rates\footnote{This can be seen from a simple parameter
counting on either sides of the seesaw equation, eq.(\ref{seesaweq}). $h^\nu$ 
contains $9$ complex parameters, $M_R$, three real whereas we only have
information about two mass squared differences and three mixing angles
in ${\mathcal M}_\nu$.}. To remedy this, either a top-down approach with specific 
SUSY-GUT models and/or flavour 
symmetries \cite{lfv2,ue3importance,lopsidedlfv,oscar} or a bottom-up 
approach with specific parameterisations of low energy unknowns have been 
adopted in the literature \cite{lfv1,petcovyag1,petcovyag2}.

\section{SO(10) and SUSY Seesaw}
In the $SO(10)$ gauge theory, all the known fermions and the 
right handed neutrinos are unified in a single representation 
of the gauge group, the \textbf{16}. 
The product of two \textbf{16} matter representations can only couple to 
\textbf{10}, \textbf{120} or \textbf{126} representations which can be 
formed either by a single Higgs field representation or a 
non-renormalisable product of representations of several Higgs fields.  
In either case, the  Yukawa matrices resulting from the couplings to 
\textbf{10} and \textbf{126} are complex symmetric whereas they are 
anti-symmetric when the couplings are to the \textbf{120}. 
Thus, the most general $SO(10)$ superpotential relevant for
fermion masses can be written as
\bea
W_{SO(10)} &=& h^{10}_{ij} 16_i~ 16_j~ 10 + h^{126}_{ij} 16_i~ 16_j~ 126 \nn \\
&+& h^{120}_{ij} 16_i~ 16_j~ 120,  
\eea
where $i,j$ refer to the generation indices. In terms of the SM fields, 
the Yukawa couplings relevant for fermion masses are given by \cite{strocchi}:
\bea
16\ 16\ 10\ &\supset& 5\ ( u u^c + \nu \nu^c) + \bar 5\
(d d^c + e e^c), \nn\\
16\ 16\ 126\ &\supset& 1\ \nu^c \nu^c + 15\ \nu \nu +
5\ ( u u^c -3~ \nu \nu^c) \nn \\ &+& \bar{45}\ (d d^c -3~ e e^c), \nn\\
16\ 16\ 120\ &\supset& 5\ \nu \nu^c + 45\ u u^c +
\bar 5\ ( d d^c + e e^c) \nn \\ &+& \bar{45}\ (d d^c -3~ e e^c),
\label{su5content}
\eea
where we have specified the corresponding $SU(5)$  Higgs representations 
for each  of the couplings and all the fermions are left handed fields. 
The resulting mass matrices can be written as
\bea
\label{upmats}
M^u &= & M^5_{10} + M^5_{126} + M^{45}_{120}, \\
\label{numats}
M^\nu_{LR} &= & M^5_{10} - 3~ M^5_{126} + M^{5}_{120}, \\
\label{downmats}
M^d &= & M^{\bar{5}}_{10} +  M^{\bar{45}}_{126} + M^{\bar{5}}_{120} 
+ M^{\bar{45}}_{120}, \\
\label{clepmats}
M^e &= & M^{\bar{5}}_{10} - 3  M^{\bar{45}}_{126} + M^{\bar{5}}_{120} 
- 3 M^{\bar{45}}_{120}, \\
M^\nu_{LL} &=& M^{15}_{126}, \\
M^\nu_{R} &=& M^{1}_{126}.
\eea

A simple analysis of the above mass matrices leads us to the following
result:  \textit{At least one
of the Yukawa couplings in $h^\nu~ =~ v_u^{-1}~M^\nu_{LR}$ has 
to be as large as the top Yukawa coupling} \cite{oscar}. This result holds 
true in general independently from
the choice of the Higgses responsible for the masses in
Eqs.~(\ref{upmats}, \ref{numats}) provided that no accidental fine
tuned cancellations of the different contributions in Eq.~(\ref{numats}) are
present. If contributions from the \textbf{10}'s solely 
dominate, $h^\nu$ and $h^u$ would be equal. If this occurs for the 
\textbf{126}'s, then $h^\nu =- 3~ h^u$. 
In case both of them have dominant
entries, barring a rather precisely fine tuned
cancellation between $M^5_{10}$ and $M^5_{126}$ in
Eq.~(\ref{numats}), we expect at least one large entry to be present in $h^\nu$.
A dominant antisymmetric contribution to top quark mass 
due to the {\bf 120} Higgs is phenomenologically excluded since it would  
lead to at least a pair of heavy degenerate up quarks. 

Apart from sharing the property that at least one eigenvalue of both $M^u$ 
and $M^\nu_{LR}$ has to be large, for the rest it is clear from 
(\ref{upmats}) and  (\ref{numats}) that these two matrices are not aligned 
in general, and hence we may expect different mixing angles appearing 
from their diagonalisation. This freedom is removed if one sticks to 
particularly simple choices of the Higgses responsible for up quark and
neutrino masses. 

We find two cases which would serve as `benchmark' scenarios for seesaw induced
lepton flavour violation in SUSY $SO(10)$. The first one corresponds
to a case where the mixing present in $h^\nu$ is small and
CKM-like. This is typical of the models where fermions attain their
masses through $10$-plets. We will call this case, `the minimal
case'. As a second case, we consider scenarios where the mixing in
$h^\nu$ is no longer small, but large like the observed PMNS
mixing. We will call this case the `the maximal case'.  

\subsection{The minimal Case: \textit{CKM mixings in $h^\nu$}}
\label{sec:CKM} 
The minimal Higgs spectrum to obtain phenomenologically viable mass matrices
includes two \textbf{10}-plets, one coupling to
the up-sector and the other to the down-sector. In this way it is possible to 
obtain the required CKM mixing \cite{buchwyler} in the quark sector. 
The $SO(10)$
superpotential is now given by

\bea
\label{primedbasis}
W_{SO(10)} &=& {1 \over 2}~  h^{u,\nu}_{ij} 16_i~ 16_j~ 10_u + 
{1 \over 2}~ h^{d,e}_{ij} 16_i ~16_j~ 10_d \nn \\ &+& 
{1 \over 2}~ h^R_{ij}~ 16_i~ 16_j~ 126. 
\eea
We further assume the {\bf 126} dimensional Higgs field gives
Majorana mass  \textit{only} to the right handed neutrinos. 
An additional feature of the above mass matrices is that all of them 
are \textit{symmetric}. 

From the above, it is clear that the following mass relations hold 
between the quark and leptonic mass matrices at the GUT 
scale\footnote{Clearly this relation cannot hold for the first two 
generations of down quarks and charged leptons.  One expects, small 
corrections due to  non-renormalisable operators or
suppressed renormalisable operators \cite{georgijarlskog} can be invoked.}: 
\be
\label{massrelations}
h^u  = h^\nu \;\;\;;\;\;\; h^d  = h^e . 
\ee
In the above basis, the symmetric $h^u$ is diagonalised by: 
\be
\label{htop}
V_{CKM}~ h^u~ V_{CKM}^{T} = h^u_{diag}. 
\ee
Hence from (\ref{massrelations}): 
\be
\label{hnumg}
 h^\nu = V_{CKM}^T~ h^u_{diag}~ V_{CKM}. \ee 
According to Eq.~(\ref{rgemi}), $\mbox{BR}(\mu \to e \gamma)$ depends on:
\be
\label{hnusqckm}
[h^\nu h^\nu]_{21} \approx h_t^2 ~V_{td}~ V_{ts} +
{\mathcal O}(h_c^2).  
\ee 
In this expression, the CKM angles are small but one would expect 
the presence of the large top Yukawa coupling to compensate such
suppression. The large couplings in $h^\nu \sim {\mathcal O}(h_t)$ induce 
significant off-diagonal entries in $m_{\tilde L}^2$ through the RG
evolution between $M_{GUT}$ and the scale of the right-handed Majorana neutrinos
\footnote{Typically one has different mass scales associated with different
right handed neutrino masses.}, $M_{R_i}$. The induced off-diagonal entries
relevant for $l_j \rightarrow l_i, \gamma$ are of the order,
\noindent 
\begin{eqnarray}
(m_{\tilde L}^2)_{21}&\approx& -{3 m_0^2+A_0^2 \over 8 \pi^2}~
h_t^2 V_{td} V_{ts} \ln{M_{GUT} \over M_{R_3}} \nn \\
\label{wcmi1}
&+& \mathcal{O}(h_c)^2, \\
(m_{\tilde L}^2)_{32}&\approx& -{3 m_0^2+A_0^2 \over 8 \pi^2}~
h_t^2 V_{tb} V_{ts} \ln{M_{GUT} \over M_{R_3}} \nn \\
\label{wcmi2}
&+& \mathcal{O}(h_c)^2, \\
(m_{\tilde L}^2)_{31}&\approx& -{3 m_0^2+A_0^2 \over 8 \pi^2}~
h_t^2 V_{tb} V_{td} \ln{M_{GUT} \over M_{R_3}} \nn \\
\label{wcmi3}
&+& \mathcal{O}(h_c)^2.
\end{eqnarray}
\noindent 
The required right handed neutrino Majorana  mass matrix consistent 
with both the observed low energy neutrino masses and mixings as well 
as with CKM like mixings in $h^\nu$ is determined easily from the 
seesaw formula defined at the scale of right handed neutrinos as 
\bea
\label{seesaw}
m_\nu &=& - h^{\nu~T}~M_R^{-1}~ h^{\nu}~v_u^2,  \\
 &=& - h^{\nu}~M_R^{-1}~ h^{\nu}~v_u^2. 
\eea
where we have used the symmetric nature of the $h^\nu$ in the second equation.
Inverting  Eq.~(\ref{seesaw}), one gets:
\bea
\label{MRstrcture1}
M_R &=&  - h^\nu~ m_\nu^{-1}~ h^\nu ~v_u^2, \nonumber \\
\label{maar2}
 &=& V_{CKM}~ h_{diag}^u~ V_{CKM}^T~ m_{\nu}^{-1} \nn \\
&& \times V_{CKM}~ h_{diag}^u~ V_{CKM}^T, 
\eea   
where we have used Eq.~(\ref{hnumg}) for $h^\nu$. Furthermore, 
$m_\nu^{-1}$ can be written as 
$m_\nu^{-1} = U_{PMNS}~ diag[m_{\nu}^{-1}]~ U_{PMNS}^T $, 
whose entries are determined at the low scale 
from neutrino oscillation experiments. The structure of $M_{R}$ can now be
derived\footnote{ The neutrino masses and mixings here are defined
at $M_{GUT}$. Radiative corrections can significantly modify the neutrino
spectrum at the weak scale \cite{nurad}. This is more true for the degenerate
spectrum of neutrino masses \cite{degenrad} and for some specific forms 
of $h^\nu$ \cite{antusch}. For our present discussion, with hierarchical
neutrino masses and up-quark like neutrino Yukawa matrices, we expect these
effects not to play a very significant role.}  
for a given set of
neutrino masses and mixing angles. Neglecting the small CKM mixing in $h^\nu$
we have
\be
\label{ckmmr}
{\ss M_R} \approx {\ss v_u^2}~ \left( \begin{array}{ccc}
{\ss h_u^2 [m^{-1}_{\nu}]_{11}} & 
\star & 
\star \\
{\ss h_u h_c [m^{-1}_{\nu}]_{12}} & 
{\ss h_c^2 [m^{-1}_{\nu}]_{22} }& 
\star \\
%h_c^2 [m^{-1}_{\nu}]_{22} & 
{\ss h_u h_t [m^{-1}_{\nu}]_{13} }&
{\ss h_c h_t [m^{-1}_{\nu}]_{23}} &
{\ss h_t^2 [m^{-1}_{\nu}]_{33} }
\end{array} \right). 
\ee
It is clear from above that the hierarchy in the $M_R$ mass matrix goes
as the square of the hierarchy in the up-type quark mass matrix.
Furthermore, for a hierarchical neutrino mass spectrum we have 
$m_{\nu_3} \approx \sqrt{\Delta m^2_{Atm}}$,
$m_{\nu_{2}} \approx \sqrt{\Delta m^2_{\odot}}$ and 
$m_{\nu_{1}} \ll \sqrt{\Delta m^2_{\odot}}$ and for a nearly bi-maximal
$U_{PMNS}$ : 
\bea
U_{PMNS} \approx \pmatrix{1/\sqrt{2} & 1/\sqrt{2} & 0 \cr
-1/2 & 1/2 &1/\sqrt{2}  \cr
1/2 & -1/2 &1/\sqrt{2}}~,~~~~
\eea
it straight-forward to check that all the right handed neutrino mass 
eigenvalues are controlled by the smallest left-handed neutrino mass. 
\be
\label{mrapproxegckm}
M_{R_3} \approx {m_t^2 \over 4~ m_{\nu_1}} \;;\;
M_{R_2} \approx {m_c^2 \over 4~ m_{\nu_1}} \;;\;
M_{R_1} \approx {m_u^2 \over 2~ m_{\nu_1}} \;.
\ee 
This implies that we can not choose an arbitrarily small neutrino mass if
we want the right-handed neutrino masses to be below $M_{GUT}$. In our 
numerical examples, we choose $m_{\nu_3} = 0.05 \mbox{ eV},
m_{\nu_2} = 0.0055 \mbox{ eV}, m_{\nu_1} = 0.001 \mbox{ eV}$.

\begin{figure}[ht]
\includegraphics[scale=0.35]{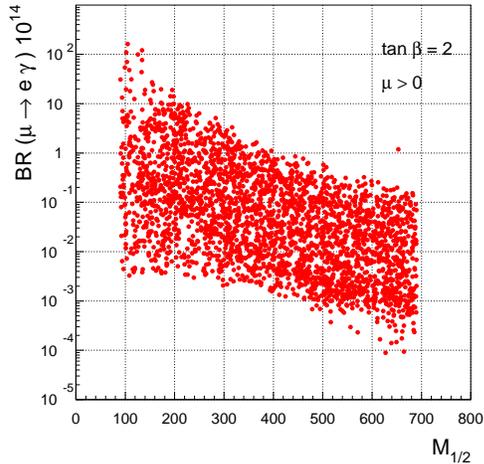}
\caption{The scatter plots of branching ratios of $\mu \to e, \gamma$
decays vs. $M_{1/2}$ are shown for the (minimal) CKM case for 
tan $\beta$ = 2. Results do not alter significantly with the change 
of sign($\mu$).}
\end{figure}

We now present numerical results for this situation in the framework
of minimal Supergravity (mSUGRA). In Fig. 1) and 2) we show the 
scatter plots for BR($\mu \rightarrow e,\gamma$ ) for the CKM case 
and $\tan \beta = 2$ and $\tan \beta = 40$ respectively. As expected 
from eq.(\ref{BR}), the BR scales with the second power in tan$\beta$. 
The plots also reflect an interesting correlation between the
branching ratios and the GUT value of the universal gaugino mass. 
This is due to the fact that the universal gaugino mass fixes the 
chargino and neutralino masses at $M_W$ and, to a small extent it 
also influences the slepton masses through RGE. However, for a fixed 
$M_{1/2}$ the different values of $m_0$ and $A_0$ can change the value 
of the BR within a range of 3 orders of magnitude. For instance, for 
$\tan \beta=40$ reaching a sensitivity of $10^{-14}$ for 
BR$(\mu \to e \gamma)$ would allow us to probe `completely' the SUSY 
spectrum up to $M_{1/2} = 300$ GeV (notice that this corresponds to 
gluino and squark masses of order 750 GeV) and would still probe a 
large regions in parameter space up to $M_{1/2} = 700$ GeV.  

\begin{figure}[ht]
\includegraphics[scale=0.35]{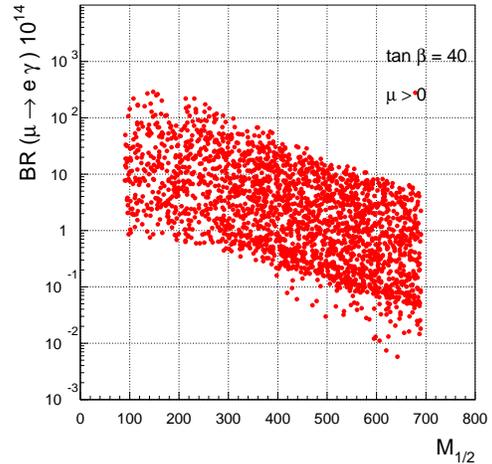}
\caption{The scatter plots of branching ratios of $\mu \to e, \gamma$
decays vs. $M_{1/2}$ are shown for the (minimal) CKM case for 
tan $\beta$ = 40. Results do not alter significantly with the change 
of sign($\mu$).}
\end{figure}

Thus in summary, though the present limits on BR($\mu \to e, \gamma)$ 
would not induce any significant constraints on the supersymmetry breaking
parameter space, an improvement in the limit to$\sim {\mathcal O }(10^{-14})$,
as being foreseen, would start imposing non-trivial constraints especially
for the large $\tan \beta$ region.

\subsection{The maximal case: \textit{PMNS mixing angles in $h^{\nu}$} }
\label{sec:PMNS}
The minimal $SO(10)$ model presented in the previous sub-section would
inevitably lead to small mixing in $h^\nu$. In fact, with two Higgs fields
in symmetric representations, giving masses to the up-sector and the 
down-sector separately, it would be difficult to avoid the small CKM like
mixing in $h^\nu$. To generate mixing angles larger than CKM angles, asymmetric
mass matrices have to be considered. In general, it is sufficient to introduce
asymmetric textures either in the up-sector or in the down-sector. In the 
present case, we assume that the down-sector couples to a combination of Higgs 
representations (symmetric and anti-symmetric) \footnote{The couplings of $\Phi$ 
in the superpotential can be either renormalisable or non-renormalisable. 
See \cite{chang} for a non-renormalisable example.}
$\Phi$, leading to an asymmetric mass matrix in the basis where the up-sector
is diagonal. As we will see below this would also require that the right handed
Majorana mass matrix to be diagonal in this basis. We have : 
\bea
\label{mnsso10}
W_{SO(10)} &=& {1 \over 2}~  h^{u,\nu}_{ii}~ 16_i ~16_i 10^u + 
{1 \over 2}~ h^{d,e}_{ij}~ 16_i ~16_j \Phi \nn \\
& + &
{1 \over 2}~ h^R_{ii}~ 16_i~ 16_i 126~, 
\eea
where the \textbf{126}, as before, generates only the right handed neutrino
mass matrix. To study the consequences
of these assumptions, we see that at the level of $SU(5)$, we have
\bea
W_{SU(5)} &=& {1 \over 2}~ h^u_{ii}~ 10_i ~10_i ~5_u 
+ h^\nu_{ii} ~\bar{5}_i~ 1_i~ 5_u \nn \\
&+ & h^d_{ij}~ 10_i ~\bar{5}_j~ \bar{5}_d 
+  {1 \over 2}~M^R_{ii}~ 1_i 1_i,
\eea
where we have decomposed the $16$ into $10 + \bar{5} + 1$ and $5_u$ and
$\bar{5}_d$ are components of $10_u$ and $\Phi$ respectively. To have large
mixing $\sim~ U_{PMNS}$  in $h^\nu$ we see that the asymmetric matrix $h^d$
should now be able to generate both the CKM mixing as well as PMNS mixing. 
This is possible if 
\be
V_{CKM}^T~ h^d~ U_{PMNS}^T = h^d_{diag}. 
\ee 
This would mean that the $10$ which contains the left handed down-quarks would
be rotated by the CKM matrix whereas the $\bar{5}$ which contains the left
handed charged leptons would  be rotated by the $U_{PMNS}$ matrix to go into
their respective mass bases \cite{chang}. Thus we have, in analogy with the
previous sub-section, the following relations hold true in the basis where 
charged leptons and down quarks are diagonal:
\bea
h^u &=& V_{CKM}~ h^u_{diag}~ V_{CKM}^T~ ,  \\
\label{hnumns}
h^\nu &=& U_{PMNS}~ h^u_{diag}. 
\eea
Using the seesaw formula of Eq.~({\ref{seesaw}) and  Eq.~(\ref{hnumns}) we have
\be
M_{R} = \mbox{Diag}\{ {m_u^2 \over m_{\nu_1}},~{m_c^2 \over m_{\nu_2}},
~{m_t^2 \over m_{\nu_3}} \}. 
\ee 
This would mean that this setup would require $M_R$ to be diagonal at the 
$SO(10)$ level in the basis of diagonal $h^{u,\nu}$, Eq.~(\ref{mnsso10}). 
We now turn our attention to  lepton flavour violation in the scenario. The
branching ratio, BR($\mu \rightarrow e, \gamma)$ would now be dependent on: 
\be
\label{hnusqmns}
[h^\nu h^{\nu~T}]_{21} = h_t^2~ U_{\mu 3}~ U_{e 3} + h_c^2~ U_{\mu 2}~ U_{e 2} +
\mathcal{O}(h_u^2). 
\ee 
It is clear from the above that in contrast to the CKM case,
the dominant contribution to the off-diagonal entries depends on the 
unknown magnitude of the element $U_{e3}$ \cite{ue3importance}. 
If $U_{e3}$ is very close to its
present limit $\sim~0.2$\cite{chooz}, the first term on the 
RHS of the Eq.~(\ref{hnusqmns})
would dominate. Moreover, this would lead to large contributions to the 
off-diagonal entries in the slepton masses with $U_{\mu 3}$ of 
${\mathcal O}(1)$. We have :
\begin{eqnarray}
(m_{\tilde L}^2)_{21}&\approx& -{3 m_0^2+A_0^2 \over 8 \pi^2}~
h_t^2 U_{e 3} U_{\mu 3} \ln{M_{GUT} \over M_{R_3}}\nn \\
\label{bcmi1}
 &+& \mathcal{O}(h_c)^2. 
\end{eqnarray}
The above contribution is large by a factor $(U_{\mu 3} U_{e3})/ (V_{td} V_{ts})
\sim 140 $ compared to the CKM case. From Eq.~(\ref{BR}) we see that 
it would mean about a factor $10^4$ times larger than the CKM case in 
BR($\mu \rightarrow e, \gamma)$. In case $U_{e3}$ is very small, \textit{i.e,} 
either zero or $\ler~ (h_c^2/h_t^2)~U_{e2}~ \sim 4 \times 10^{-5}$, the 
second term $\propto ~h_c^2$
in Eq.~(\ref{hnusqmns}) would dominate. However the off-diagonal contribution
in slepton masses, now being proportional to charm Yukawa could be much smaller, 
in fact, even  smaller than the CKM contribution by a factor 
\be{h_c^2~ U_{\mu 2} ~U_{e 2} \over  h_t^2 ~V_{td} ~V_{ts}} 
\sim  7 \times 10^{-2}.
\ee
If $U_{e3}$ is close to it's present limit, the current bound on
BR($\mu \rightarrow e, \gamma$) would already be sufficient to
produce stringent limits on the SUSY mass spectrum. Similar
$U_{e3}$ dependence can be expected in the $\tau \to e $ transitions
where the off-diagonal entries are given by : 
\begin{eqnarray}
(m_{\tilde L}^2)_{31}&\approx& -{3 m_0^2+A_0^2 \over 8 \pi^2}~
h_t^2 U_{e 3} U_{\tau 3} \ln{M_{GUT} \over M_{R_3}} \nn \\
\label{bcmi3}
&+& \mathcal{O}(h_c)^2. 
\end{eqnarray}
The $\tau \to \mu$ transitions are instead $U_{e3}$-independent probes of SUSY,
whose importance was first pointed out in Ref.~\cite{kingblazektmg}. As in the
rest of the cases, the off-diagonal entry in this case is given by :
\begin{eqnarray}
(m_{\tilde L}^2)_{32}&\approx& -{3 m_0^2+A_0^2 \over 8 \pi^2}~
h_t^2 U_{\mu 3} U_{\tau 3} \ln{M_{GUT} \over M_{R_3}} \nn \\
\label{bcmi2}
&+& \mathcal{O}(h_c)^2. 
\end{eqnarray}

\begin{figure}[ht]
\includegraphics[scale=0.35]{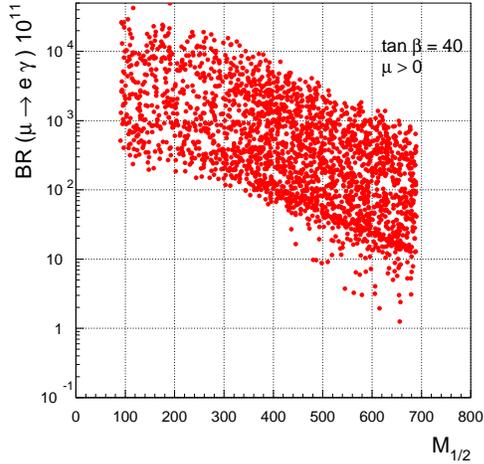}
\caption{The scatter plots of branching ratios of $\mu \to e, \gamma$
decays vs. $M_{1/2}$ are shown for the (maximal) PMNS case for 
tan $\beta$ = 40. Results do not alter significantly with the change 
of sign($\mu$).}
\end{figure}

In the PMNS scenario Fig. 3) shows the plot for BR($\mu
\rightarrow e, \gamma$) for tan $\beta$ = 40.  As we said in the
earlier, in the PMNS case, the results concerning 
BR($\mu \rightarrow e, \gamma$) strongly depend on
the unknown value of $U_{e3}$. In this plot, the value of
$U_{e3}$ chosen is very close to the present experimental upper limit
\cite{chooz}. As long as $U_{e3} \ger 4 \times 10^{-5}$, the plots
scale as $U_{e3}^2$, while for $U_{e3} \ler 4 \times 10^{-5}$ the term
proportional to $m_c^2$ in Eq.~(\ref{bcmi1}) starts dominating
and then, the result is insensitive to the choice of $U_{e3}$.  
For instance, a value of $U_{e3} =0.01$ would reduce the BR by a factor of
225 and still a significant amount of  the parameter space 
for $\tan \beta = 40$ would
be excluded. 
We further find that with the present limit on BR($\mu \rightarrow e, \gamma$),
all the parameter space would be completely excluded up to $M_{1/2}=
300$ GeV for $U_{e3} =0.15$, for any vale of $\tan \beta$. 

In the $\tau \to \mu \gamma$ decay the situation is similarly constrained.
For $\tan \beta=2$, the present bound of $3 \times 10^{-7}$ starts probing
the parameter space up to $M_{1/2} \leq 150$ GeV. The main difference is 
that this does not depend on the value of $U_{e3}$, and therefore
it is already a very important constraint on the parameter space of 
the model.  In fact, for large $\tan \beta = 40$, as shown in Fig.(4), 
reaching the expected limit of $6 \times 10^{-8}$ would be
able to rule out completely this scenario up to gaugino masses of $400$ GeV 
and only a small portion of the parameter space with heavier gauginos 
would survive. In the limit $U_{e3} =0$, this decay mode would provide
a stronger constraint on the model, than $\mu \to e, \gamma$ which would
now be suppressed as it contains only contributions proportional to $h_c^2$,
as shown in eq.(\ref{bcmi1}). 

\begin{figure}[ht]
\includegraphics[scale=0.35]{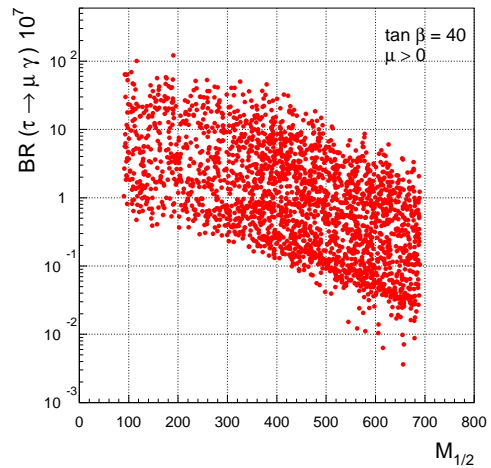}
\caption{The scatter plots of branching ratios of $\tau \to \mu, \gamma$
decays vs. $M_{1/2}$ are shown for the (maximal) PMNS case for 
tan $\beta$ = 40. Results do not alter significantly with the change 
of sign($\mu$).}
\end{figure}

In summary, in the PMNS/maximal mixing case, even the present limits
from BR($\mu \to e, \gamma$ ) can rule out large portions of the supersymmetric
breaking parameter space, if $U_{e3}$ is either close to its present limit
or within an order of magnitude of it (as soon, the planned experiments
might find out \cite{goodman}). These are more severe for the large $\tan \beta$
case. In the extreme situation of $U_{e3}$ being zero or very small $\sim 
{\mathcal O}(10^{-4} - 10^{-5})$, BR($\tau \to \mu, \gamma$) will start
playing an important role, with its present constraints already disallowing
large regions of the parameter space at large $\tan \beta$. 

\subsection{Correlations with other SUSY Search Strategies}
In addition to the improvements in LFV experiments, this is also 
going to be decade where we should be able to establish whether
low energy supersymmetry exists or not through direct searches
at Large Hadron Collider (LHC) \cite{lhc}. On the other hand, 
improved astrophysical observations from experiments like 
WMAP\cite{wmap} and Planck are going to determine the relic density 
of supersymmetric LSP at unprecedented accuracy. Within mSUGRA 
correlations between these two search strategies have been studied
\cite{baertata}. Incorporating the seesaw mechanism in the model 
\`a la SO(10), would generate another discovery strategy through
the lepton flavour violation channel.  This is especially
true when the LFV entries in the slepton mass matrices are maximised,
as in the PMNS case. 

We see that three main regions in the mSUGRA parameter space would 
survive after imposing
all the present phenomenological and astrophysical 
constraints\cite{profumo}\footnote{For a bottom-up analyses,
 see Ref.\cite{campbell}.}.
These are: (a). The stau coannihilation regions, where lightest stau
is quasi-degenerate with the neutralino LSP and efficient stau-stau as well
as stau-neutralino (co)-annihilations suppress the relic density. (b). The
A-pole funnel region, where the neutralino(bino)-neutralino annihilation
process is greatly enhanced through a resonant s-channel exchange of the heavy
neutral Higges A and H and (c). Focus point or Hyperbolic branch regions, where
non-negligible higgsino fraction in the lightest neutralino is produced. 
In each of these regions the LFV rates emanating from the seesaw mechanism 
can be computed and contrasted with the sensitivity of direct searches at 
LHC. Assuming the maximal mixing PMNS case, we find \cite{profumo}:

\begin{itemize}

\item \textit{Coannihilation Regions}: In these regions, which are
mostly accessible at LHC, an improvement of two orders of magnitude in 
the branching ratio sensitivity from the present limit, 
would make $\mu \to e \gamma$ visible 
for most of the parameter space as long as $U_{e3} \gtrsim 0.02$, even for 
the low tan $\beta$ region. For large tan $\beta$, independent of $U_{e3}$,
$\tau \to \mu \gamma$ will start probing this region provided a sensitivity
of $\mathcal{O}(10^{-8})$ is reached.  

\item \textit{A-pole funnel Regions}: In these regions the LHC
reach is not complete and LFV may be competitive. If $U_{e3} \gtrsim 10^{-2}$,
the future $\mu \to e \gamma$ experiments, with limits of 
${\mathcal O}(10^{-14})$ will probe most of the parameter 
space. 
As before, $\tau \to \mu \gamma$ will probe this region once the BR 
sensitivity reaches $\mathcal{O}(10^{-8})$.

\item \textit{Focus Point Regions}: Since the LHC reach in this region is 
rather limited due to the
large $m_0$ and $M_{1/2}$ values, LFV could constitute a privileged road 
towards SUSY discovery. This would require improvements of at least a couple
of orders of magnitude (or more, depending on the value of $U_{e3}$) of
improvement on the present limit of BR($\mu \to e, \gamma$). 
DM searches will also have in future partial access 
to this region, leading 
to a new complementarity between LFV and the quest for the cold dark matter
 constituent of the universe.
\end{itemize}

\section{Seesaw induced Hadronic FCNC}
So far we have seen that the SUSY version of the seesaw mechanism can 
lead to potentially large leptonic flavour violations, so much that they
 could even compete with the direct searches like LHC. If one combines 
these ideas of supersymmetric seesaw with those of quark-lepton unification,
 as in a supersymmetric Grand Unified Theory (GUT), one would expect that
 the seesaw resultant flavour effects now would also be felt in the
 hadronic sector and vice-versa \cite{hkr,bhs}. In fact, this is what
 happens in a SUSY SU(5) with seesaw mechanism \cite{moroi}, where the
 seesaw induced  RGE effects generate flavour violating terms in 
the right handed squark multiplets. 
However, as is the case with MSSM + seesaw mechanism, within the SU(5)
model also, information from the neutrino masses is not sufficient to 
fix all the seesaw parameters; a large neutrino Yukawa coupling has
to be \textit{assumed} to have the relevant phenomenological consequences
in hadronic physics, like CP violation in $B \to \Phi K_s$ etc. 

As we have already seen within the $SO(10)$ model, a large neutrino
Yukawa, of the order of that of the  top quark, is almost inevitable. 
Using this, it has been pointed in Ref.\cite{chang}, that the observed
large atmospheric $\nu_\mu ~-~\nu_\tau$ transitions imply a potentially
large $b \to s$ transitions in SUSY SO(10). In the presence of CP
violating phases this can lead to enhanced CP asymmetries in $B_s$ and $B_d$
decays. In particular, the still controversial discrepancy between the 
SM prediction and the observed $A_{CP}(B_d \to \Phi K_s)$\cite{bellebabar}
 can be attributed
to these effects. Interestingly enough, despite the severe constraints on the
$b \to s$ transitions from $B \to X_s,\gamma$ \cite{bsgamma,borzumati},
 subsequent detailed analyses
\cite{otherbs,rome} proved that there is still enough room for sizable
deviations from the SM expectations for CP violation in the $B$ systems. 
The readers interested in various correlations in $b \to s$ transitions
with all possible FV off-diagonal squark mass entries can find an exhaustive
answer in Ref.\cite{rome}. 

Finally let us make a short comment about possible correlations between 
the hadronic and leptonic FV effects in a SUSY-GUT. If the FV soft 
breaking terms appear at a scale larger than that of the Grand Unification, 
then they must be related by the GUT symmetry. This puts constraints on the
boundary conditions for the running of the FV soft parameters. From this
consideration, one might intuitively expect that some correlation between 
various leptonic and hadronic FCNC processes \cite{ourprl} can occur at the weak 
scale. If in the evolution of the sparticle masses from the Grand Unification
scale down to the electroweak scale, one encounters the seesaw physics, then
the quark-lepton correlations involving the left-handed sleptons, though 
modified, lead to even stronger constraints on hadronic
 physics \cite{ourprl,workinprogress}.

\section{Conclusions}
Undoubtedly, the seesaw mechanism represents (one of ) the best 
proposals to generate naturally small neutrino masses. But, how
can we make sure that this is indeed the Nature's choice ? 
Even establishing the Majorana nature of the neutrinos through a
positive evidence of neutrinoless double beta decay, it will be
difficult to assess that such Majorana  masses come from a seesaw. 
Indeed, as we said at the beginning, in the SM seesaw we expect 
very tiny charged LFV effects, probably without any chance to ever
observe them. When moving to SUSY seesaw we add an important handle
to our effort to establish the presence of a seesaw. In fact, as we
tried to show in this talk, SUSY extensions of the SM with a seesaw
have a general ``tendency" to enhance (or even strongly enhance) rare
LFV processes. Hence the combination of the observation of neutrinoless
double beta decay and of some charged LFV phenomenon would constitute
an important clue for the assessment of SUSY seesaw in Nature.

There is no doubt that after the discovery of the neutrino masses, 
among the indirect tests of SUSY through FCNC and CP violating 
phenomena, LFV processes have acquired a position of utmost relevance. 
It would be spectacular, if by the time LHC observes the first SUSY
particle, we could see also a muon decaying to an electron and a photon ! 
After thirty years, we could have the simultaneous confirmation of two
of the most challenging physics ideas: seesaw and low energy SUSY. 

\textbf{Acknowledgements:} We would like to thank all our collaborators,
without whom the results presented in this talk would not have been 
possible.

\end{document}